\documentclass[11pt,a4paper]{article}
\pdfoutput=1
\usepackage{jheppub}
\usepackage{graphicx}
\usepackage{lipsum}
\usepackage[all]{hypcap}

\newcommand{\beq}{\begin{equation}}
\newcommand{\eeq}{\end{equation}}
\newcommand{\pol}{{\cal P}}

\begin{document}

\title{Measuring $\boldsymbol{c}$-quark polarization in $\boldsymbol{W}$+$\boldsymbol{c}$ samples\\at ATLAS and CMS}

\affiliation{Department of Particle Physics and Astrophysics\\ 
		     Weizmann Institute of Science\\
		     Rehovot 7610001, Israel}
\author[1]{Yevgeny Kats\note{Present address: Theoretical Physics Department, CERN, CH-1211 Geneva 23, Switzerland. E-mail: \href{mailto:yevgeny.kats@cern.ch}{\tt yevgeny.kats@cern.ch}.}}
\emailAdd{yevgeny.kats@weizmann.ac.il}

\abstract{The process $pp \to W^-c$ produces polarized charm quarks. The polarization is expected to be partly retained in $\Lambda_c$ baryons when those form in the $c$-quark hadronization. We argue that it will likely be possible for ATLAS and CMS to measure the $\Lambda_c$ polarization in the $W$+$c$ samples in Run~2 of the LHC. This can become the first measurement ever of a longitudinal polarization of charm quarks. Its results will provide a unique input to the understanding of polarization transfer in fragmentation. They will also allow applying the same measurement technique to other (e.g., new physics) samples of charm quarks in which the polarization is a priori unknown. The proposed analysis is similar to the ATLAS and CMS measurements of the $W$+$c$ cross section in the 7 TeV run that used reconstructed $D$-meson decays for charm tagging.}

\maketitle

\section{Introduction}

\begin{figure}[t]
\begin{center}
\includegraphics[width=0.8\textwidth]{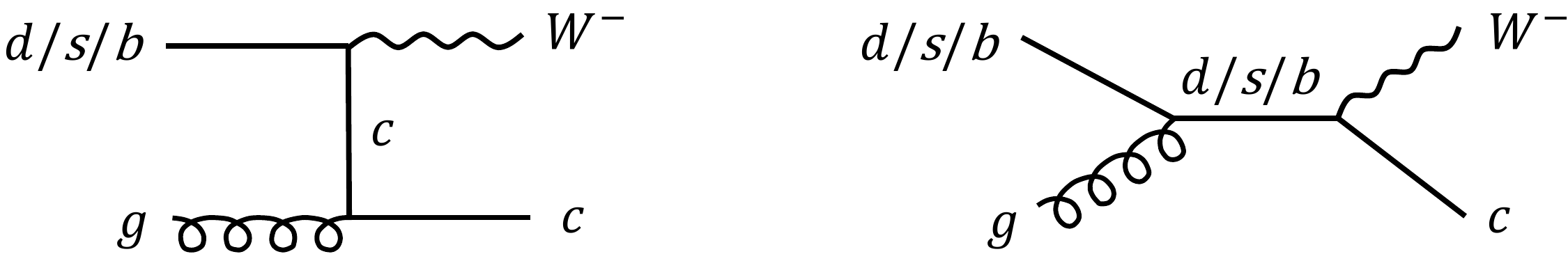}
\end{center}
\caption{Leading-order diagrams for $W$+$c$ production. The contribution with the $s$ quark in the initial state is an order-of-magnitude larger than the CKM-suppressed $d$-quark contribution, while the $b$-quark contribution is entirely negligible.}
\label{fig:diagrams}
\end{figure}

The process
\beq
p p \to W^- c
\label{pp-Wc}
\eeq
proceeds at leading order via the diagrams in figure~\ref{fig:diagrams} (see also~\cite{Giele:1995kr,Stirling:2012vh}).\footnote{Throughout the paper, the conjugate process $p p \to W^+ \overline c$ is included implicitly. Its cross section is slightly lower than that of $p p \to W^- c$, primarily because the valence quarks of the proton include a $d$ but not a $\bar d$.} Since the $W$ couples only to the left-handed quark fields, the charm quarks in the final state are polarized. In this paper we argue that ATLAS and CMS can likely measure this polarization in Run~2 of the LHC.

While measurements of top-quark polarization are now standard~\cite{Khachatryan:2015dzz,Chatrchyan:2013wua,Khachatryan:2016xws,Aad:2013ksa,Abazov:2012oxa,Abazov:2015fna,Abazov:2016tba}, measuring the polarization of all the other quarks is more challenging because they hadronize. It is, however, possible. For a heavy quark, $m_q \gg \Lambda_{\rm QCD}$, an ${\cal O}(1)$ fraction of the quark polarization is expected to be retained when its hadronization produces a baryon, which is most commonly a $\Lambda_b$ in the $b$-quark case or a $\Lambda_c$ in the $c$-quark case~\cite{Mannel:1991bs,Ball:1992fw,Falk:1993rf,Galanti:2015pqa}. The polarization can then be determined from kinematic distributions of the baryon decay products. Evidence of $\Lambda_b$ polarization has been seen in $e^+e^- \to Z \to b\bar b$ events at LEP~\cite{Buskulic:1995mf,Abbiendi:1998uz,Abreu:1999gf}. There have been no analogous measurements for the $\Lambda_c$. Even though the charm quark is not as heavy as the bottom, it is reasonable to expect the $\Lambda_c$ to also carry polarization, since the LEP experiments observed ${\cal O}(1)$ polarization retention even for the (strange-based) $\Lambda$ baryons~\cite{Buskulic:1996vb,ALEPH:1997an,Ackerstaff:1997nh}.

Carrying out charm-quark polarization measurements in Standard Model samples at the LHC is important for several reasons. First, establishing the measurement technique on a Standard Model sample will allow applying it confidently to any new-physics sample that will contain charm quarks. The polarization can in turn provide key information about the new-physics Lagrangian. Importantly, as alluded to above, the $c$-quark polarization fraction carried by the $\Lambda_c$ is currently unknown. Therefore, for a full interpretation of any new-physics measurement, it would be useful to have this fraction extracted from a known sample, such as $W$+$c$. Second, regardless of whether any new physics is discovered at the LHC, the polarization measurements will advance our understanding of fragmentation. There are various theoretical approaches that attempt to describe the polarization transfer from a heavy quark to a baryon. Some parameterize the nonperturbative QCD physics in terms of certain quantities that can be determined also by other types of measurements~\cite{Falk:1993rf} (see also~\cite{Galanti:2015pqa}), as we will review, and others attempt to model it~\cite{Chow:1996df,Adamov:2000is}. These frameworks will greatly benefit from experimental inputs.

The possibility of using $\Lambda_c$ decays for measuring charm-quark polarization in ATLAS and CMS was first analyzed in~\cite{Galanti:2015pqa}. It was shown there, that in Run~2 it would be possible to do such measurements in $pp \to t\bar t$ samples, where polarized charm quarks are produced in $W^+\to c\bar s$ decays. The process in eq.~\eqref{pp-Wc} might be even more promising than $t\bar t$ because it provides an order-of-magnitude larger sample of charm quarks (although the background is also larger).

A relatively clean sample of $W$+$c$ events can be obtained by using the decays $W \to \ell\nu$ (where $\ell = e$ or $\mu$) along with charm tagging. Such samples were used by ATLAS~\cite{Aad:2014xca} and CMS~\cite{Chatrchyan:2013uja} in the 7~TeV run for measuring the $W$+$c$ cross section. Similar measurements were done by CDF~\cite{Aaltonen:2007dm,Aaltonen:2012wn,Aaltonen:2015aka} and D0~\cite{Abazov:2008qz,Abazov:2014fka} at the Tevatron and by LHCb~\cite{Aaij:2015cha}. The tagging of charm quarks in these analyses was based on either a reconstructed multiprong $D^+$ or $D^{\ast+}$ decay, or an inclusively defined displaced decay, or a soft muon from a semileptonic decay. In the polarization measurement, one would instead use reconstructed decays of charmed baryons, most importantly the $\Lambda_c^+$, for tagging the charm and measuring its polarization. While ATLAS and CMS have not yet reported any $W$+$c$ analyses from the 8 or 13~TeV runs, measurements of inclusive $W$ production at 13~TeV~\cite{Aad:2016naf,CMS-PAS-SMP-15-004} show that the ATLAS and CMS triggers still allow collecting the $W \to \ell\nu$ decays without a significant decrease in efficiency. It is therefore realistic to obtain large samples of $W$+$c$ events in Run~2.

Measuring the $\Lambda_c$ polarization in $W$+$c$ samples may also be possible at LHCb. While LHCb's luminosity and acceptance (and therefore statistics) are inferior relative to ATLAS and CMS, its particle identification, vertexing ability and momentum resolution offer significant advantages from the point of view of purity. This makes it plausible for an LHCb measurement to be competitive with ATLAS and CMS~\cite{MikeWilliams}. However, it will not be possible for us to analyze a potential LHCb measurement in this work, as this would require dedicated simulation tools, unlike in the case of ATLAS/CMS where we will be able to build upon an existing analysis.

Note that inclusive QCD production is less attractive than $W$+$c$ for charm polarization measurements. In QCD production, $pp\to c\bar c$, the quark polarization~\cite{Dharmaratna:1996xd} is small because it only appears at the next-to-leading order in $\alpha_s$ and is further suppressed at high $p_T$ by the quark mass. Additionally, the strong dependence of the polarization on the parton-level kinematics of the event complicates the interpretation. Moreover, since this polarization is transverse, it is not protected from being generated by soft QCD effects~\cite{Mulders:1995dh,Anselmino:2000vs}. Differently, longitudinal polarization, as expected in $W$+$c$ production, is protected by the parity invariance of QCD and can therefore arise only due to the polarization of the charm quarks from the hard process.

That being said, measurements of the $\Lambda_c$ transverse polarization in QCD samples, which we will not discuss here, are interesting in their own right, regardless of whether they will be more sensitive to the polarization of the quarks from the hard process or soft QCD effects. Such measurements have already been done with very soft $\Lambda_c$'s in the fixed-target experiments BIS-2~\cite{Aleev:1984bd}, R608~\cite{Chauvat:1987kb}, NA32 (ACCMOR)~\cite{Jezabek:1992ke} and E791~\cite{Aitala:1999uq}. The results showed a very large transverse polarization, for which an interpretation attempt was made in~\cite{Goldstein:1999jr}.

\section{Overview of the theoretical picture}
\label{sec:theoretical}

When a polarized $c$ quark produces a $\Lambda_c$, a large fraction of its polarization is expected to be preserved in the $\Lambda_c$ polarization. The basic reason is that in the heavy quark limit, $m_c \gg \Lambda_{\rm QCD}$, the $\Lambda_c$ can be viewed as a bound state of the original $c$ quark and a spin-0 diquark~\cite{Falk:1993rf}.

The likely dominant polarization loss effect is due to events in which the $c$ quark first hadronizes to a $\Sigma_c$ or $\Sigma_c^\ast$ (also known as $\Sigma_c(2455)$ and $\Sigma_c(2520)$, respectively) and then a $\Lambda_c$ is produced via $\Sigma_c^{(\ast)} \to \Lambda_c\pi$~\cite{Falk:1993rf}.\footnote{We conservatively assume that since the pions in $\Sigma_c^{(\ast)} \to \Lambda_c\pi$ are quite soft, such events cannot be separated out very effectively by ATLAS and CMS.} In the heavy quark picture, the $\Sigma_c$ (spin $1/2$) and $\Sigma_c^\ast$ (spin $3/2$) are the states obtained from combining the spin-$1/2$ $c$ quark with a spin-$1$ light diquark:
\beq
\frac12 \otimes 1 = \frac12 \oplus \frac32
\eeq
\beq
\quad c \quad qq' \quad \Sigma_c \;\;\, \Sigma_c^\ast \;.\nonumber
\eeq
Therefore, when a $c$ quark in a definite spin state hadronizes with a spin-1 diquark, it in general produces a superposition of $\Sigma_c$ and $\Sigma_c^\ast$ states. However, the subsequent decay $\Sigma_c^{(\ast)} \to \Lambda_c \pi$ occurs in either the $\Sigma_c$ or $\Sigma_c^\ast$ mass eigenstate, neither of which is a $c$-quark spin eigenstate. As a result, the polarization gets reduced. This is different from the case of a direct hadronization into a $\Lambda_c$, where the light diquark has spin 0, and then the polarization of the $\Lambda_c$ is the polarization of the original $c$ quark:
\beq
\frac12 \otimes 0 = \frac12
\eeq
\beq
\quad c \quad\, qq' \;\;\; \Lambda_c \;. \nonumber
\eeq
Using the appropriate Clebsch-Gordan coefficients, and taking the case of a longitudinal $c$-quark polarization as relevant to $W$+$c$ production, one obtains that for $\Lambda_c$'s produced via $\Sigma_c^{(\ast)}$'s the polarization is reduced by the factor~\cite{Falk:1993rf,Galanti:2015pqa}
\beq
\left.\frac{\pol(\Lambda_c)}{\pol(c)}\right|_{\Sigma_c^{(\ast)}\, {\rm production}} = \frac19 + \frac49\,w_1 \,,
\eeq
where $w_1$, to be discussed below, is the probability for the spin-1 diquark to form with a spin component $1$ or $-1$ (rather than $0$) with respect to the fragmentation axis~\cite{Falk:1993rf}. When the $\Sigma_c^{(\ast)}$ widths are taken into account~\cite{Galanti:2015pqa}, the result becomes
\beq
\left.\frac{\pol(\Lambda_c)}{\pol(c)}\right|_{\Sigma_c^{(\ast)}\, {\rm production}} \approx 0.07 + 0.46\,w_1 \,.
\eeq
The longitudinal polarization retention fraction of the full $\Lambda_c$ sample is then given by
\beq
r_L \equiv \frac{\pol(\Lambda_c)}{\pol(c)} \approx \frac{1 + A\left(0.07 + 0.46\,w_1\right)}{1+A} \,,
\label{rL}
\eeq
where $A$ is the ratio of the rates of $\Lambda_c$ production from $\Sigma_c^{(\ast)}$ decays and direct $\Lambda_c$ production, which is related to the formation probability of a spin-1 vs.\ spin-0 diquark~\cite{Falk:1993rf}.

The values of $A$ and $w_1$ are determined by nonperturbative QCD physics, so there is no easy way for computing them reliably. However, they can be measured. In particular, $A$ can be determined by measuring the $\Sigma_c^{(\ast)}$ yields, and $w_1$ by measuring the angular distributions of the pions in the $\Sigma_c^{(\ast)} \to \Lambda_c\pi$ decays~\cite{Falk:1993rf,Galanti:2015pqa}. The CLEO experiment found~\cite{Brandenburg:1996jc}
\beq
w_1 = 0.71 \pm 0.13 \,,
\label{w1}
\eeq
a value consistent with the isotropic limit, $w_1 = 2/3$. An explanation for this has been suggested in~\cite{Chow:1996df}. However, this result is in some tension with what was obtained for the analogous $b$-quark system by the DELPHI experiment at LEP, $w_1 = -0.36 \pm 0.30 \pm 0.30$~\cite{DELPHI-95-107,Feindt:1995qm,Podobrin:1996yu}, and with the phenomenological model of~\cite{Adamov:2000is} which predicts $w_1 \approx 0.39$. The value of $A$ can be estimated~\cite{Galanti:2015pqa} from the relative $\Sigma_c/\Lambda_c$ yield reported by the E791 experiment~\cite{Aitala:1996cy} to be\footnote{Here we took only the experimental uncertainty into account. It is difficult to estimate the theoretical uncertainty associated with extrapolating (using the approach of~\cite{Galanti:2015pqa}) the numbers obtained for $\Sigma_c^0$ and $\Sigma_c^{++}$ to the other $\Sigma_c^{(\ast)}$ states.}
\beq
A = 1.1 \pm 0.4 \,.
\label{A}
\eeq
This is smaller than the expectation from the naive statistical hadronization model, $A \approx 2.6$ (see~\cite{Galanti:2015pqa} for details), and the prediction of the phenomenological model of~\cite{Adamov:2000is}, $A \approx 6$.

Tentatively accepting the results in eqs.~\eqref{w1} and~\eqref{A}, we obtain from eq.~\eqref{rL} that
\beq
r_L = 0.68 \pm 0.06\,.
\eeq
Taking into account that one-loop QCD corrections to the hard process reduce the $c$-quark polarization by about $3\%$~\cite{Korner:1993dy}, one gets the estimate $\pol(\Lambda_c) \approx -0.66$, where the negative sign indicates that the polarization is left-handed. We hope that LHCb and/or the $B$ factories will also measure $w_1$ and $A$, to make this prediction more robust. These measurements do not require polarized charm quarks, and can therefore be done even in samples of $pp\to c\bar c$ or $e^+e^- \to c\bar c$. LHCb has already released one analysis involving $\Lambda_c$'s~\cite{Aaij:2013mga}, and Belle has even done high-precision studies of $\Sigma_c^{(\ast)}$'s~\cite{Lee:2014htd}, but no results relevant to $A$ or $w_1$ have been reported yet.

So far, we have not discussed any possible direct effects on the charm-quark spin from the nonperturbative QCD processes that govern the initial, QCD-scale, stage of the hadronization. In the heavy quark limit, $m_c \gg \Lambda_{\rm QCD}$, such effects are suppressed because the chromomagnetic moment, through which the QCD would interact with the spin, is proportional to $1/m_c$. But in reality $\Lambda_{\rm QCD}/m_c \sim 0.2$, so ${\cal O}(\Lambda_{\rm QCD}/m_c)$ corrections may be nonnegligible. However, not much is known about them. The measurements we propose in this paper will help to assess the size of this effect.

Finally, we would like to note that the $\Lambda_c$ polarization will in general have some dependence on the fraction of the $c$-quark momentum carried by the $\Lambda_c$, $z$. It would be interesting to measure this dependence, also known as the polarized (or spin-dependent) fragmentation function (FF), once a sufficient amount of data is available. For the $\Lambda$ baryon, the polarization has already been measured as a function of $z$ in the hadronic $Z$ decays at LEP~\cite{Buskulic:1996vb,ALEPH:1997an,Ackerstaff:1997nh}. (See also~\cite{deFlorian:1997zj} for a theoretical interpretation in terms of the FFs of the different contributing quark flavors.) For the $\Lambda_c$, only the unpolarized FF, which describes the $z$-dependent probability for a $c$ quark to produce a $\Lambda_c$, has been measured~\cite{Avery:1990bc,Seuster:2005tr,Aubert:2006cp}. A prediction of the polarized FFs for the $\Lambda_c$, relying on a particular model, was made in~\cite{Adamov:2000is}. Polarized FFs are instrumental in computing the renormalization group evolution of the polarization retention with the scale of the hard process~\cite{Stratmann:1996hn} (as exemplified in~\cite{Adamov:2000is}), which has been neglected in the above discussion.

\section{\texorpdfstring
{$\boldsymbol{\Lambda_c}$ reconstruction}
{Lambda-c reconstruction}}
\label{sec:reconstruction}

A convenient decay mode for reconstructing the $\Lambda_c^+$, with a relatively large branching fraction, ${\cal B} = (6.84^{+0.32}_{-0.40})\%$~\cite{PDG}, is
\beq
\Lambda_c^+ \to p K^- \pi^+ .
\eeq
Notice that its signature is similar to that of
\beq
D^+ \to K^- \pi^+ \pi^+ ,
\eeq
a decay mode that was used in the ATLAS~\cite{Aad:2014xca} and CMS~\cite{Chatrchyan:2013uja} $W$+$c$ measurements in the 7~TeV run. Both decays are characterized by one negative and two positive tracks emerging from a displaced vertex and passing through the entire tracker. The track momenta reconstruct the known hadron mass when the tracks are assigned the correct particle identities.

\begin{figure}[t]
\begin{center}
\includegraphics[width=0.49\textwidth]{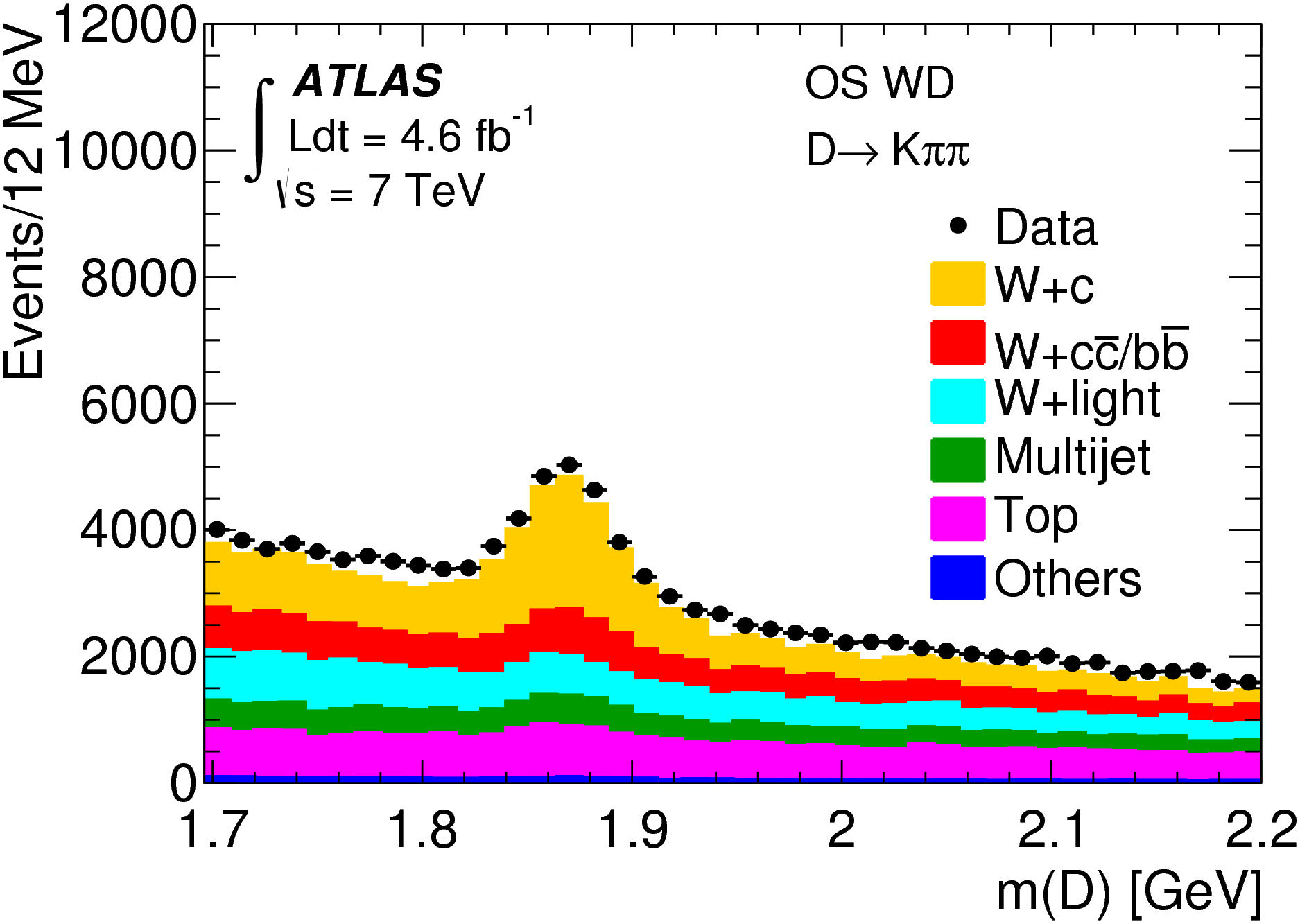}
\includegraphics[width=0.49\textwidth]{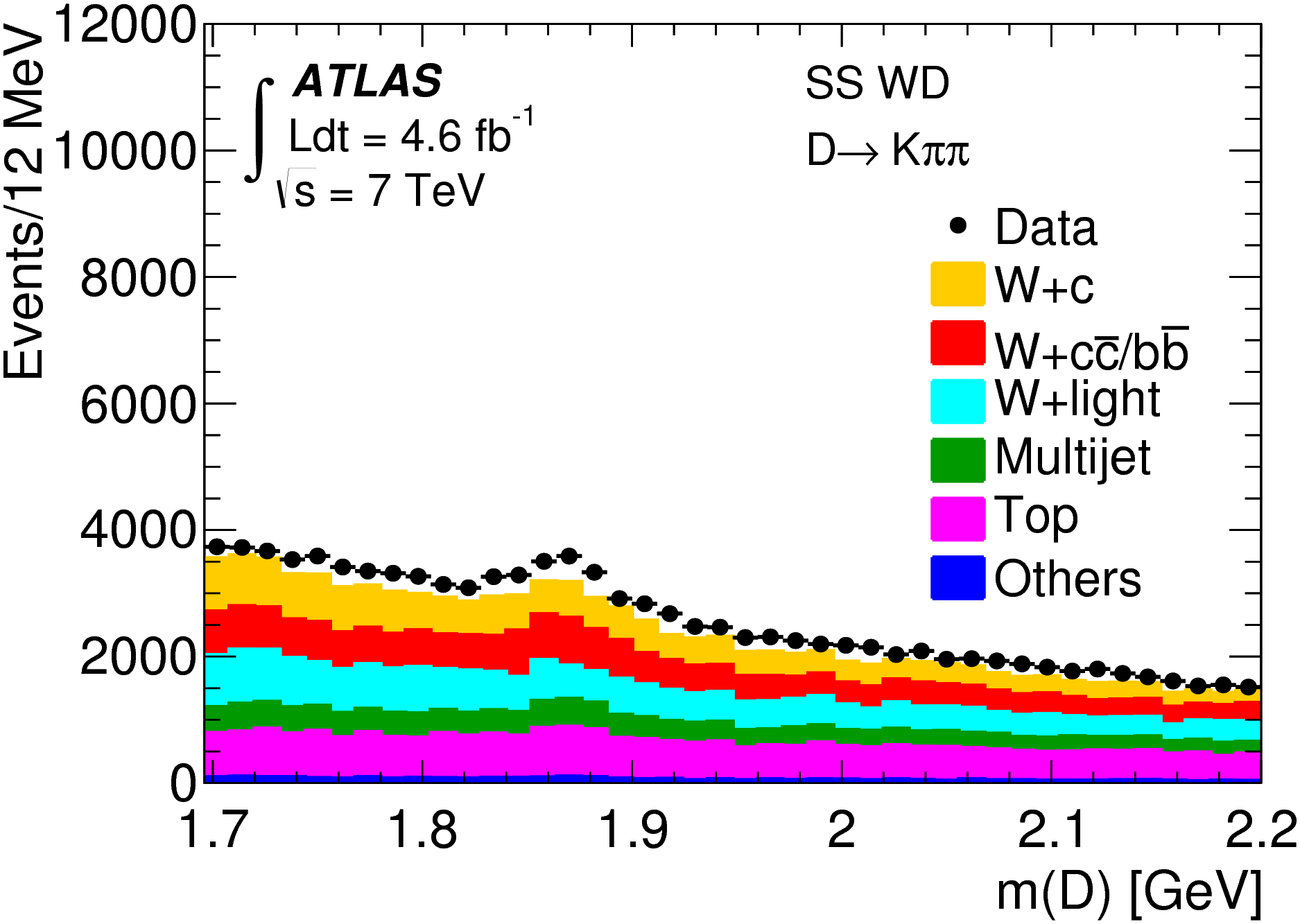}\\\vskip 2mm
\includegraphics[width=0.49\textwidth]{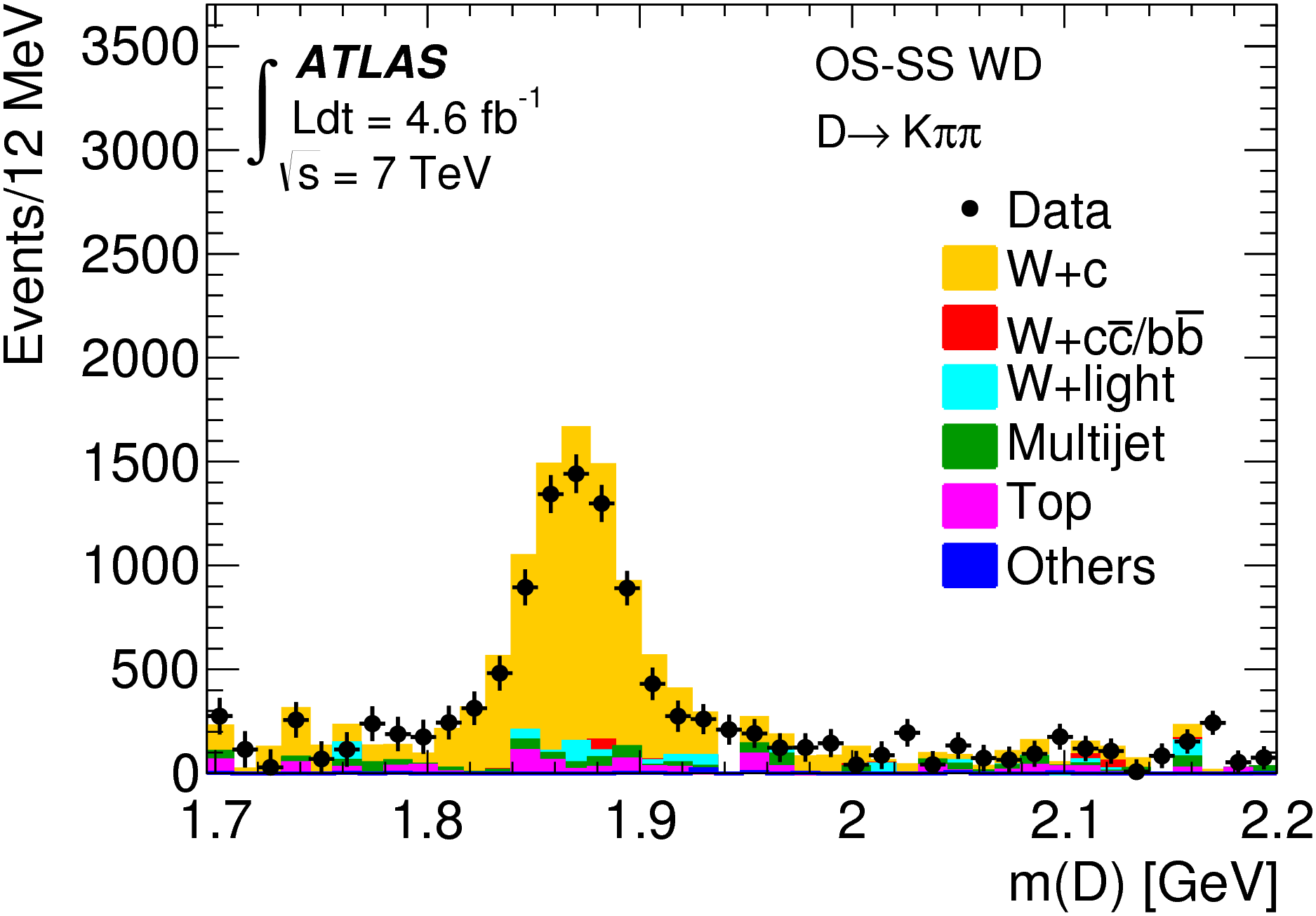}\qquad
\includegraphics[width=0.45\textwidth]{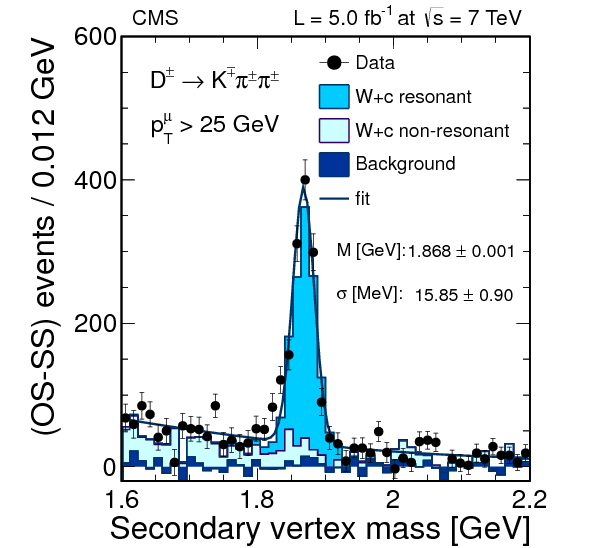}
\vspace{-4mm}
\end{center}
\caption{Reconstructed $D^\pm \to K^\mp \pi^\pm \pi^\pm$ decays in the $W$+$c$ samples of the 7~TeV LHC run. First row: opposite-sign (left) and same-sign (right) event yields from ATLAS~\cite{Aad:2014xca}.\protect\footnotemark\ The backgrounds were obtained by ATLAS from simulation, except for the multijet background which was estimated using a control region in data. Second row: Differences between the opposite- and same-sign event yields from ATLAS~\cite{Aad:2014xca}\textsuperscript{\ref{footnote-aux-mat}} (left) and CMS~\cite{Chatrchyan:2013uja} (right).}
\label{fig:D-mass}
\end{figure}

Figure~\ref{fig:D-mass}, top-left, shows the prominent peak obtained by ATLAS in the reconstructed $D^\pm$ mass distribution. Figure~\ref{fig:D-mass}, bottom row, shows the even much cleaner peaks obtained by both ATLAS and CMS when taking the difference between the event yields of candidates with the right and wrong $D^\pm$ charge, with respect to what is expected based on the charge of the lepton from the $W$. The subtraction is very effective because for many of the backgrounds the $D^\pm$ candidate is equally likely to have either charge.

\addtocounter{footnote}{-1}
\footnotetext{Auxiliary material for ref.~\cite{Aad:2014xca}, STDM-2012-14, is available at \url{https://atlas.web.cern.ch/Atlas/GROUPS/PHYSICS/PAPERS/STDM-2012-14/}.\label{footnote-aux-mat}}

$\Lambda_c^+$ reconstruction can be done along the same lines as the $D^+$ reconstruction. It is more difficult for two reasons. First, for a given collider energy and integrated luminosity, the $\Lambda_c^+$ peak will be reduced relative to the $D^+$ peak by the factor
\beq
\frac{f(c\to D^+)\,{\cal B}(D^+\to K^-\pi^+\pi^+)}{f(c\to\Lambda_c^+)\,{\cal B}(\Lambda_c^+\to pK^-\pi^+)} \approx 5.3 \,,
\label{f-BR-ratio}
\eeq
where we used the fragmentation fractions from~\cite{Lisovyi:2015uqa}. In particular, $f(c\to\Lambda_c^+) = (6.23 \pm 0.41)\%$. At the same time, the smooth background (as in figure~\ref{fig:D-mass}, top-left) will remain roughly the same. The potential doubling of the smooth background due to the ambiguity in assigning candidate particle identities to tracks ($p$ vs.\ $\pi^+$) can be avoided by noticing that in a decay of an energetic $\Lambda_c$, the proton almost always carries more momentum than the pion (in the lab frame), due to kinematics. We also assume that the contribution of $D^+ \to K^- \pi^+ \pi^+$ to the background is easily suppressed by requiring inconsistency with the $D^+$ hypothesis.

In Run~2, background fluctuations will go down because of an increase in statistics by a factor of $\sim 60$ (due to a factor of $\sim 3$ increase in the $W$+$c$ cross section and a factor of $\sim 20$ increase in the integrated luminosity). The signal-to-background ratio will not be affected much by the transition from 7 to 13~TeV because the cross sections of the signal and background processes increase by similar factors.\footnote{We use the NLO cross sections from {\sc MadGraph5\_aMC@NLO}~\cite{Alwall:2014hca} with the cuts $p_T > 20$~GeV, $|\eta|<2.5$ on the $c$ quark (or the light parton, for the $W$+light background) and no cuts on the $W$. The CMS analysis~\cite{Chatrchyan:2013uja} required $p_T^{c{\rm-jet}} > 25$~GeV, and the ATLAS analysis~\cite{Aad:2014xca} $p_T^{D^+} > 8$~GeV, which corresponds to $p_T^{c{\rm-jet}} \gtrsim 16$~GeV (since a charmed hadron carries $\sim 50\%$ of the $c$-jet momentum on average~\cite{Barate:1999bg,Chatrchyan:2013uja}). We find the 13~TeV/7~TeV cross section ratios to be $2.8$ for $W$+$c$, $2.1$ for $W$+light jets, and $2.4$ for $W$+$c\bar c$ and $W$+$b\bar b$. For $t\bar t$, the corresponding ratio of the inclusive NNLO+NNLL cross sections~\cite{Czakon:2013goa,Czakon:2011xx} is 4.8.}

The second difficulty in the $\Lambda_c^+ \to pK^-\pi^+$ reconstruction is that the displacement of the decay vertex, which was used by both ATLAS and CMS in the $D^+ \to K^- \pi^+ \pi^+$ selection, is less pronounced for the $\Lambda_c^+$ due to its shorter lifetime,
\beq
\tau_{\Lambda_c^+} \approx \frac{\tau_{D^0}}{2} \approx \frac{\tau_{D_s^+}}{2.5} \approx \frac{\tau_{D^+}}{5} \,.
\label{lifetimes}
\eeq
In the CMS analysis, for example, less than 20\% of the charm events (consisting of approximately 61\% $D^0$, 24\% $D^+$, 8\% $D_s^+$ and $6\%$ $\Lambda_c^+$~\cite{Lisovyi:2015uqa}) had a well-identified secondary vertex~\cite{Chatrchyan:2013uja}. For the full reconstruction of $D^+\to K^-\pi^+\pi^+$, CMS had an efficiency of about 11\% (for $p_T^{c{\rm-jet}} > 25$~GeV)~\cite{Chatrchyan:2013uja} and ATLAS reported an efficiency of 32\% (for $p_T^{D^+} > 8$~GeV)~\cite{Aad:2014xca}. Relaxing the vertex requirements for $\Lambda_c^+$ selection would increase backgrounds involving prompt tracks.

However, this should be somewhat less of an issue in Run~2, since vertexing performance will improve due to the ``Insertable $B$-Layer'' (IBL) that has been installed in the ATLAS detector~\cite{ATLAS-TDR-19,ATLAS-TDR-19-ADD,Aad:2012wf,ATL-PHYS-PUB-2015-018}, and the upcoming upgrade of the CMS pixel detector~\cite{CMS:2012sda}.\footnote{ATLAS's IBL is a fourth pixel layer, which is closer to the beam axis than the innermost pixel layer in Run~1 (3.3~cm instead of 5.0~cm) and has smaller pixels ($50~\mu\mbox{m} \times 250~\mu\mbox{m}$ rather than $50~\mu\mbox{m} \times 400~\mu\mbox{m}$). In CMS, the innermost pixel layer remains at 4.4~cm, with a pixel size of $100~\mu\mbox{m} \times 150~\mu\mbox{m}$. The CMS pixel detector is planned to be upgraded in the winter of 2016-2017. Its innermost layer will then be at 3.0~cm while the pixel size will remain the same.} These improvements will also help to cope with the increased pileup, whose potential effects were not taken into account in the above discussion.

Furthermore, as was noted in~\cite{Galanti:2015pqa}, the lifetime differences, eq.~\eqref{lifetimes}, may actually be useful for reducing $D$-meson backgrounds. This would be analogous to what ATLAS did in~\cite{Aad:2014nra,Aad:2015gna}, where lifetime differences between the bottom and charmed hadrons (e.g., $\tau_{B^+} \approx \tau_{B^0} \approx 1.5 \tau_{D^+} \approx 4 \tau_{D^0}$) were one of the handles for tagging $c$ jets while rejecting $b$ jets. Additionally, the large difference between the $\Lambda_c$ and bottom-hadron lifetimes, $\tau_{\Lambda_c} \approx \tau_b/7$, can be used for suppressing backgrounds in which $\Lambda_c$'s are produced in $b$ jets.

\section{\texorpdfstring
{$\boldsymbol{\Lambda_c}$ polarization measurement}
{Lambda-c polarization measurement}}
\label{sec:pol-measurement}

The $\Lambda_c$ polarization, $\pol(\Lambda_c)$, can be measured using the angular distribution of any of the $\Lambda_c^+ \to p K^- \pi^+$ decay products. In the $\Lambda_c^+$ rest frame, the distributions are
\beq
\frac{1}{\Gamma}\frac{d\,\Gamma}{d \cos \theta_i} = \frac12\left(1 + \alpha_i\pol(\Lambda_c)\cos\theta_i\right) ,
\label{angular}
\eeq
where $\theta_i$ describes the direction of motion of the product $i = p, K^-, \pi^+$ relative to the polarization direction, and $\alpha_i$ are the so-called spin analyzing powers (or asymmetry parameters). The same distributions describe the decays of $\overline\Lambda_c^-$'s of an opposite polarization. It should be noted that if the selection efficiency strongly depends on the energies of the decay products, unfolding needs to be applied, since eq.~\eqref{angular} assumes inclusiveness in the energies.

By analyzing transverse polarization data from the NA32 experiment (ACCMOR), it was found~\cite{Jezabek:1992ke} that $|\alpha_{K^-}| \gg |\alpha_p|,\, |\alpha_{\pi^+}|$, i.e., the angular distribution of the kaon is much more sensitive to the $\Lambda_c$ polarization than those of the proton and the pion. Furthermore, for one of their samples, $\alpha_{K^-}\pol(\Lambda_c) = -0.65^{+0.22}_{-0.18}$\,. Since, by definition, $|\pol(\Lambda_c)| \leq 1$ and $|\alpha_i| \leq 1$, this indicates that $|\alpha_{K^-}|$ is ${\cal O}(1)$, consistent with the theoretical conjecture in~\cite{Bjorken:1988ya}. The sign of $\alpha_{K^-}$ is conjectured to be negative~\cite{Bjorken:1988ya}, as for $\alpha_\Lambda$ in $\Lambda_c^+ \to \Lambda\ell^+\nu_\ell$ and $\Lambda_c^+ \to \Lambda\pi^+$, where experiments indeed find $\alpha_\Lambda = -0.86 \pm 0.04$ and $-0.91 \pm 0.15$, respectively~\cite{PDG}. This is also the sign that can be inferred from the NA32 results by assuming that the direction of the transverse $\Lambda_c$ polarization is the same as that for similarly produced $\Lambda$'s~\cite{Jezabek:1992ke,Barlag:1994bk}.

The decay $\Lambda_c^+ \to p K^-\pi^+$ has a nonresonant as well as several resonant channels ($p\overline K^\ast(892)^0$, $\Delta(1232)^{++} K^-$ and $\Lambda(1520)\pi^+$)~\cite{PDG,Aitala:1999uq}, without much interference between them~\cite{Aitala:1999uq}. Predictions for the spin analyzing properties of two of the resonant channels were made in~\cite{Konig:1993wz} using the quark-model approach of~\cite{Korner:1992wi}. For the $p\overline K^\ast(892)^0$ channel, which is responsible for $(21\pm 3)\%$ of the rate~\cite{Aitala:1999uq,PDG}, they find $\alpha_p \approx 0.69$, with a possibly large theoretical uncertainty. Correspondingly, $\alpha_{\overline K^\ast(892)^0} \approx -0.69$, likely leading to a sizable negative value for $\alpha_{K^-}$ in this channel, consistent with the above discussion. For the $\Delta(1232)^{++}K^-$ channel, responsible for $(17 \pm 4)\%$ of the rate~\cite{Aitala:1999uq,PDG}, the same approach predicts $\alpha_p, \alpha_{K^-}, \alpha_{\pi^+} \approx 0$~\cite{Konig:1993wz}. The nonresonant channel constitutes $(55 \pm 6)\%$ of the rate~\cite{Aitala:1999uq,PDG}.

Overall, since the precise values of $\alpha_i$ are unknown, the proposed analysis will in practice measure the products $\alpha_i\pol(\Lambda_c)$. This does not present an obstacle to using the $W$+$c$ measurement for calibrating a future measurement in a new-physics sample, since the $\alpha_i$ will be the same prefactors in the two cases.

\section{Sensitivity estimate for Run 2}

Because of the large dependence of the $\Lambda_c^+ \to pK^-\pi^+$ selection and reconstruction on vertexing, it is difficult for us to fully simulate the proposed analysis. Instead, to obtain a rough idea about the attainable $\Lambda_c^+ \to pK^-\pi^+$ reconstruction efficiency and purity, we will use the information provided on $D^+ \to K^-\pi^+\pi^+$ reconstruction in the ATLAS $W$+$c$ analysis~\cite{Aad:2014xca}.

\begin{figure}[t]
\begin{center}
\includegraphics[width=0.42\textwidth]{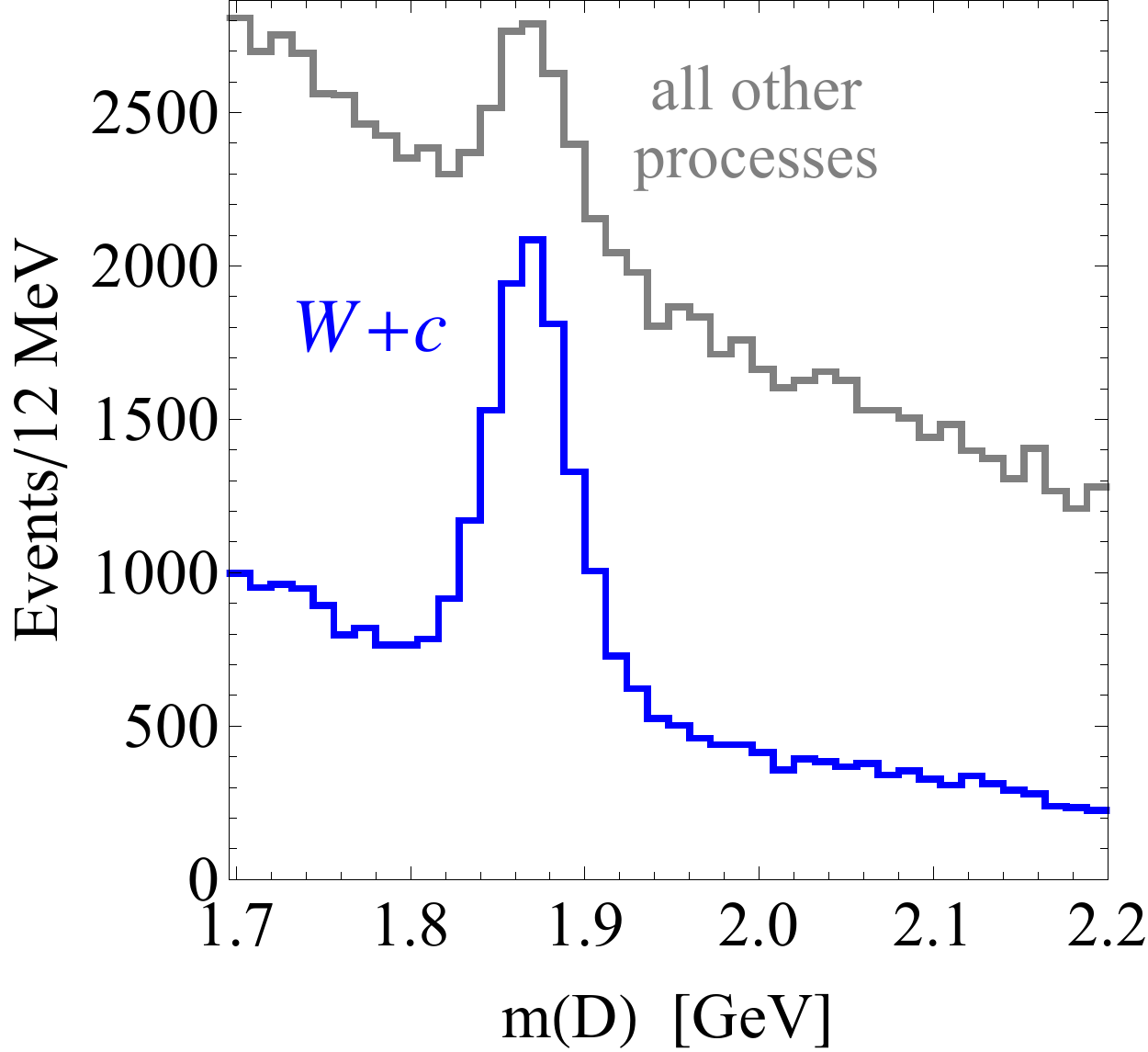}\qquad
\includegraphics[width=0.42\textwidth]{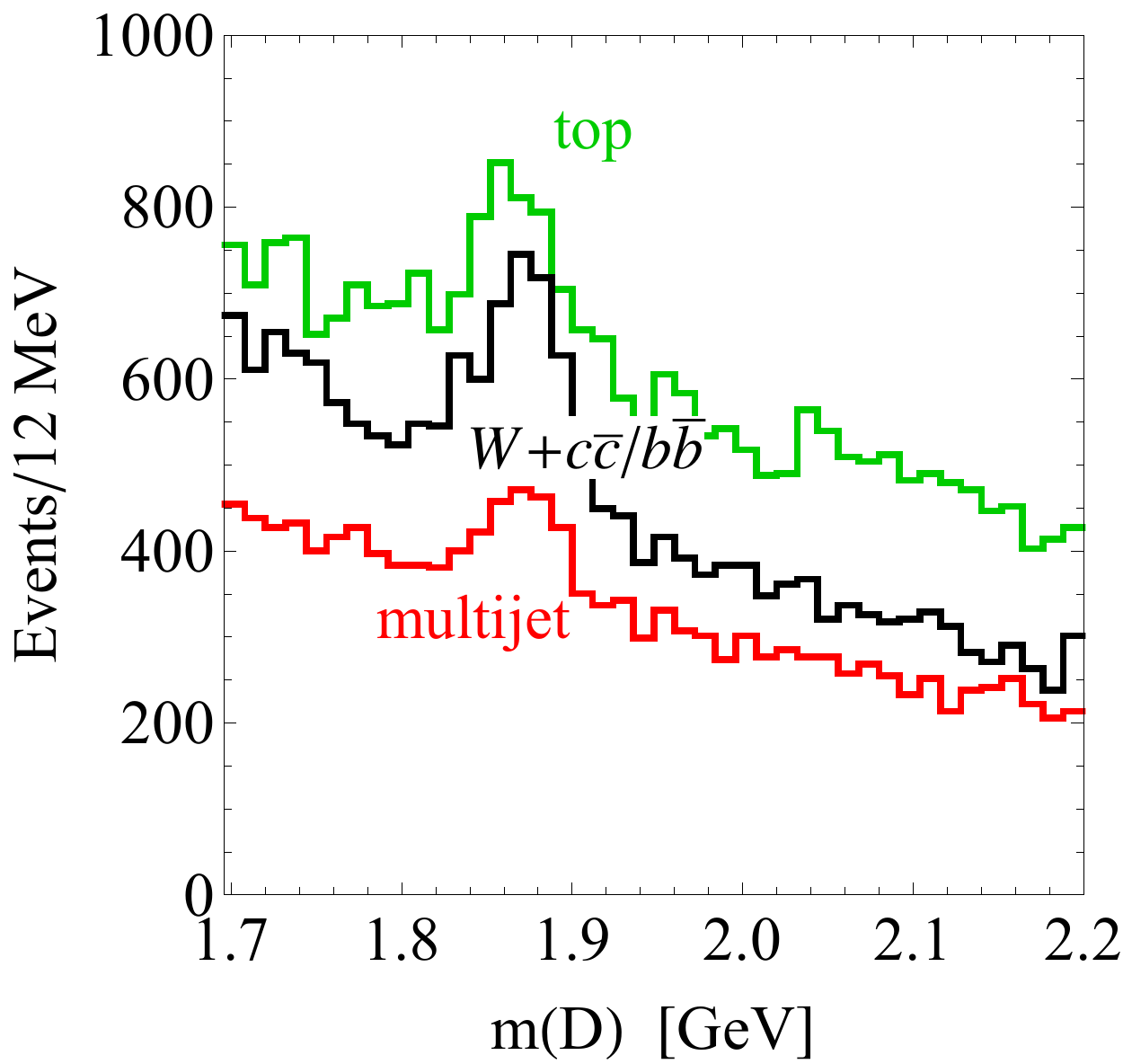}
\vspace{-4mm}
\end{center}
\caption{Separate (unstacked) contributions to the ATLAS plot for $D^\pm \to K^\mp \pi^\pm \pi^\pm$ (our figure~\ref{fig:D-mass}, top-left).
Left: $W$+$c$ (blue) and all the other processes (gray). Right: top processes (green), $W$+$c\bar c/b\bar b$ (black) and multijet processes (red).}
\label{fig:D-mass-separate}
\end{figure}

In figure~\ref{fig:D-mass}, top-left, we already showed ATLAS's estimates for the various backgrounds. As shown more clearly in figure~\ref{fig:D-mass-separate}, left, in addition to the desired peak, the $W$+$c$ process itself contributes a smooth background. It arises due to a variety of situations in which the event is found to contain a displaced 3-prong vertex not due to $D^+ \to K^-\pi^+\pi^+$. As can also be seen in figure~\ref{fig:D-mass-separate}, left, the contribution from processes other than $W$+$c$ likewise includes both a smooth component and a peak at the $D^+$ mass. The peak is due to real $D^+ \to K^-\pi^+\pi^+$ decays present mainly in the processes shown in figure~\ref{fig:D-mass-separate}, right. In these processes, the size of the peak relative to the smooth component is smaller than for $W$+$c$ due to additional smooth contributions from non-$c$ jets. We expect the backgrounds for $\Lambda_c^+ \to pK^-\pi^+$ to behave qualitatively similarly.

In the $\Lambda_c$ analysis we propose here, in addition to determining the size of the backgrounds, one will need to know their contribution to the polarization. For the peaking parts of the backgrounds, it will be the polarization of actual $\Lambda_c$'s present in the event, while for the smooth parts, it will be the mostly-fake polarization obtained from interpreting the 3-prong vertex as if it originated from a $\Lambda_c^+ \to pK^-\pi^+$ decay. (It is not entirely fake. For example, the small contribution due to the 4-body decay $\Lambda_c^+ \to pK^-\pi^+\pi^0$ will have some dependence on the $\Lambda_c^+$ polarization.) It is also important to note that part of the $\Lambda_c$'s in the background events will arise not from hadronization of charm quarks but from decays of bottom baryons and will generally be polarized even if the bottom baryons are not. Let us now discuss the different background contributions in more detail.

In the multijet processes, the charm quarks, and therefore also the $\Lambda_c$'s they produce, are unpolarized, because of the parity invariance of QCD. The charm quarks from $W^+\to c\bar s$ decays in the top-quark backgrounds (which include $t\bar t$ and single-top processes) are polarized in the same way as those of the $W$+$c$ events. The resulting $\Lambda_c^+$ decays can therefore be considered as a small additional contribution to the signal, after accounting for the sign of the accompanying lepton. Bottom quarks in top, $W$+$b\bar b$ and multijet events contribute $\Lambda_c$'s with a different polarization, due to the electroweak decays of the bottom hadrons (primarily the $\Lambda_b$) to $\Lambda_c$'s and $\Sigma_c^{(\ast)}$'s. Since $\tau_b \approx 7\tau_{\Lambda_c}$, this contribution can probably be estimated by using a control sample enriched in $\Lambda_c$'s produced at a highly displaced vertex. Finally, the contribution from $W$+$c\bar c$ processes will cancel in the difference between the polarization of $\Lambda_c$'s in the right-sign samples and that of $\overline\Lambda_c$'s in the wrong-sign samples, analogous to figure~\ref{fig:D-mass}, bottom row, because the $g\to c\bar c$ vertex is CP conserving.

Backgrounds due to events other than $\Lambda_c^+ \to pK^-\pi^+$ can likely be estimated using sidebands. Their contribution to the reconstructed polarization will vary smoothly with the reconstructed mass, as long as the known value of the $\Lambda_c^+$ mass is not being used in determining the $\Lambda_c^+$ rest frame for the polarization measurement. An exception will be a small and narrow polarized peak due to $\Xi_c^+ \to pK^-\pi^+$ near $m_{\Xi_c^+} \approx 2468$~MeV~\cite{Jun:1999gn,Link:2001rn}.

To estimate the statistical precision of the polarization measurement, let us consider for simplicity the kaon forward-backward asymmetry with respect to the $\Lambda_c$ flight direction, in the $\Lambda_c$ rest frame, as the variable sensitive to the kaon angular distribution in eq.~\eqref{angular}:
\beq
{\cal A}_{\rm FB}
\equiv \frac{N(\cos\theta_{K^-} > 0) - N(\cos\theta_{K^-} < 0)}{N} \,,
\eeq
where $N = N(\cos\theta_{K^-} > 0) + N(\cos\theta_{K^-} < 0)$ is the total number of events reconstructed in the $\Lambda_c$ mass window. The signal contribution is
\beq
{\cal A}_{\rm FB,S} = \frac{\alpha_{K^-}\pol(\Lambda_c)\,S}{2N} \,,
\eeq
where $S$ is the number of signal events, which include only correctly reconstructed $\Lambda_c^+ \to p K^-\pi^+$ decays from $W$+$c$ production. The statistical uncertainty on ${\cal A}_{\rm FB}$ is given by
\beq
\sigma({\cal A}_{\rm FB}) = \sqrt{\frac{1 - {\cal A}_{\rm FB}^2}{N}}
\simeq \frac{1}{\sqrt N} \,.
\eeq
Since the $c$-quark polarization satisfies
\beq
\pol(c) \propto \pol(\Lambda_c) \propto {\cal A}_{\rm FB,S} \,,
\eeq
the expected statistical significance for the observation of a non-zero $\pol(c)$ is
\beq
\frac{|{\cal A}_{\rm FB,S}|}{\sigma({\cal A}_{\rm FB})}
= \frac{|\alpha_{K^-}\pol(\Lambda_c)|}{2}\frac{S}{\sqrt N}
\label{precision}
\eeq
standard deviations. The relative statistical precision of the measurement is the inverse of this quantity.

To obtain a ballpark figure, let us take the example of the ATLAS $D^+$ peak (figure~\ref{fig:D-mass-separate}, left)\footnote{We fit the backgrounds to a second-order polynomial after excluding the mass range $1870 \pm 102$~MeV. We define the $D^+$ mass window, which contains about $95\%$ of the peak, as $1870 \pm 42$~MeV. The peak width, for both $D^+$ and $\Lambda_c^+$, is dominated by detector resolution effects. As we only seek a rough estimate, we neglect the difference between the $D^+$ and $\Lambda_c^+$ masses ($m_{\Lambda_c^+} \approx 2286$~MeV), the different masses of their decay products which affect both the peak resolution (via the distributions of the decay products momenta) and the shape of the smooth backgrounds (which are reconstructed by assigning the expected particle identities to the three tracks), and differences between the ATLAS and CMS resolutions.} and account, as discussed in section~\ref{sec:reconstruction}, for the difference between the $D^+ \to K^-\pi^+\pi^+$ and $\Lambda_c^+ \to p K^- \pi^+$ event rates, and the higher statistics ($\approx 100$~fb$^{-1}$) anticipated for Run~2. Let us also assume, that to cancel some of the background contributions, the analysis will use the difference between the opposite- and same-sign samples. The cost of this, in terms of the statistical fluctuations of the background, is approximately a doubling of $N$. We then get $S/\sqrt N \approx 47$. The value of $|\alpha_{K^-}\pol(\Lambda_c)|$ that enters eq.~\eqref{precision} will be measured in the analysis itself. As a reasonable possibility, in line with the discussions of the expected $\Lambda_c$ polarization in section~\ref{sec:theoretical} and the spin analyzing power of the kaon in section~\ref{sec:pol-measurement}, let us take $|\alpha_{K^-}\pol(\Lambda_c)| = 0.4$. We then obtain a precision of $11\%$.

The estimate in the previous paragraph is optimistic since it does not take into account the expected degradation in the efficiency and/or purity due to the shorter $\Lambda_c^+$ lifetime, whose impact is difficult for us to estimate. This issue will be only partly mitigated by the Run~2 pixel detector upgrades mentioned in section~\ref{sec:reconstruction}. However, even if relaxed requirements on the vertex displacement let more background in, for instance increasing $N$ by a factor of~2, and still admit less signal $S$, for example also by a factor of~2, eq.~\eqref{precision} still predicts high statistical significance, of about $3\sigma$. We also note that the ATLAS analysis used only very basic properties of the event in the selection procedure. It is therefore plausible that backgrounds can be significantly reduced further by targeting the remaining non-$W$+$c$ contributions in a more dedicated way.

\section{Comments on other possible strategies}

While we have focused on $\Lambda_c^+ \to pK^-\pi^+$, other reconstructible decay modes of the $\Lambda_c^+$ can potentially be used for the polarization measurement as well. The decay $\Lambda_c^+\to\Lambda\pi^+$ is attractive because of its known and large spin analyzing power, $\alpha_\Lambda = -0.91 \pm 0.15$~\cite{PDG}. However, its branching fraction is not very large, ${\cal B} \approx 0.9\%$ after requiring $\Lambda\to p\pi^-$ (cf.\ ${\cal B} \approx 7\%$ for $\Lambda_c^+ \to p K^-\pi^+$), and the $\Lambda$ reconstruction efficiency in ATLAS and CMS is only ${\cal O}(10\%)$~\cite{Aad:2011hd,ATLAS:2014ona,Khachatryan:2011tm} because the $\Lambda$ decays far from the interaction point. The BIS-2 experiment~\cite{Aleev:1984bd} used the decay $\Lambda_c^+ \to p\overline K^0\pi^+\pi^-$ (${\cal B} \approx 1.2\%$ after requiring $\overline K^0 \to \pi^+\pi^-$), where it found some evidence for sizable spin analyzing powers for the $p$ and $\overline K^0$. Both the BIS-2~\cite{Aleev:1984bd} and R608~\cite{Chauvat:1987kb} used the decay $\Lambda_c^+ \to \Lambda\pi^+\pi^+\pi^-$ (${\cal B} \approx 2.2\%$ after requiring $\Lambda \to p\pi^-$), with the R608 finding evidence for a large spin analyzing power for the $\Lambda$.

We note, in addition, that the spin analyzing powers of $\Lambda_c^+ \to p K^-\pi^+$ can be obtained by analyzing this decay in conjunction with $\Lambda_c^+\to\Lambda\pi^+$ in the high-statistics samples of $\Lambda_c$'s from the decays of inclusively-produced $b$ hadrons, in particular at LHCb. These $\Lambda_c$'s are expected to be highly polarized due to the electroweak nature of the $b\to c$ transition. (For theoretical studies of the polarization in some of the $\Lambda_b\to\Lambda_c$ decays, see~\cite{Korner:1991ph,Bialas:1992ny,Konig:1993wz,Konig:1993ze,Lee:1998bj,Cardarelli:1998tq,Albertus:2004wj,Ebert:2006rp,Gutsche:2015mxa}.) The known spin analyzing power of $\Lambda_c^+\to\Lambda\pi^+$ enables measuring this polarization, making it possible to extract the spin analyzing powers of $\Lambda_c^+ \to p K^-\pi^+$.

It is also possible to measure the $\Lambda_c$ polarization by doing an amplitude analysis of the $\Lambda_c^+ \to p K^-\pi^+$ decay, using the various contributions with intermediate resonances, as was done by the E791 experiment~\cite{Aitala:1999uq}. We did not discuss this method here only for simplicity.

\section{Discussion}

With Run~2 of the LHC, we enter an era in which ATLAS and CMS can measure not only the momentum but also the spin state (polarization) of quarks produced in hard processes. It has already been pointed out that the polarizations of $b$ and $c$~\cite{Galanti:2015pqa} as well as $s$ quarks~\cite{Kats:2015cna} can be measured in the Standard Model $t\bar t$ samples of Run~2. It has also been demonstrated~\cite{Kats:2015cna} that in the future, quark polarization measurements in certain new physics scenarios can be both feasible and useful, after they have been calibrated on known Standard Model samples. In the current paper, we studied the possibility of measuring the $c$-quark polarization in Standard Model $W$+$c$ events by utilizing the $\Lambda_c^+ \to pK^-\pi^+$ decays. Using rough estimates, we found such an analysis to likely be doable in Run~2.

The $W$+$c$ samples offer an advantage over the $t\bar t$ samples considered in~\cite{Galanti:2015pqa} from the point of view of the signal statistics. The cross section of
\beq
pp \to W^-c\,,\quad
W^- \to \ell^-\bar\nu
\eeq
is larger than that of
\beq
pp \to t\bar t\,,\quad
t \to (W^+\to c\bar s)\,b\,,\quad
\bar t \to (W^- \to \ell^-\bar\nu)\,\bar b
\eeq
by a factor of about 13 (and similarly for the conjugate processes).\footnote{For this estimate, for $W$+$c$ production, we used the NLO cross section from {\sc MadGraph5\_aMC@NLO}~\cite{Alwall:2014hca} with the cuts $p_T > 20$~GeV, $|\eta|<2.5$ on the $c$ quark. For $t\bar t$ production, we used the NNLO+NNLL cross section~\cite{Czakon:2013goa,Czakon:2011xx} without any kinematic cuts.} On the other hand, the $W$+$c$ samples suffer from sizable backgrounds from several other processes (figures~\ref{fig:D-mass} and~\ref{fig:D-mass-separate}), while in single-lepton $t\bar t$ samples, contaminations from non-$t\bar t$ processes are small (see, e.g.,~\cite{ATLAS-CONF-2015-049,CMS:2015toa}); contaminations from light jets in $t\bar t$ events cannot be large either. Therefore, as far as purity is concerned, $t\bar t$ samples seem advantageous over $W$+$c$. Even though in the ATLAS $D^+$ analysis the total background in the mass window was larger than just the intrinsic $W$+$c$ background by only a factor of about 5 (see figure~\ref{fig:D-mass-separate}, left), it will likely be yet larger in the $\Lambda_c^+$ analysis. This is because accommodating the smaller displacement of the $\Lambda_c^+$ decay vertex will increase the contributions from non-$W$+$c$ backgrounds that involve prompt tracks, even though the recent (ATLAS) and upcoming (CMS) pixel detector upgrades will help somewhat. It is therefore not obvious a priori which of the two measurements, $t\bar t$ or $W$+$c$, will end up being more powerful. Both are likely useful. We have also mentioned the possibility of a measurement in the $W$+$c$ sample of LHCb, which will have lower statistics but higher purity.

The proposed measurements will provide information about the longitudinal polarization retention in $\Lambda_c$ baryons in charm-quark fragmentation. Remarkably, this effect has not yet been measured in any experiment. These measurements, as well as those proposed for the bottom~\cite{Galanti:2015pqa} and strange~\cite{Kats:2015cna} quarks, will provide precious inputs to our understanding of the spin dynamics in the fragmentation process. The charm quark is unique in this regard, in that it is not as heavy relative to the QCD scale as the bottom quark, and not in the entirely nonperturbative regime like the strange and the lighter quarks. From a completely different perspective, the charm polarization measurements in the Standard Model samples will prepare the ground for using similar measurements to figure out the nature of new physics processes that produce charm quarks, once such processes are discovered.

\acknowledgments
I am grateful to Andrea Giammanco, Jan Kretzschmar, Alex Pearce, Jonathan Shlomi, Emmanuel Stamou and Mike Williams for their comments on a draft of this paper.

\bibliographystyle{utphys}
\bibliography{Wc}

\end{document}